\documentclass[twocolumn]{aastex63}
\shorttitle{Magnetar giant flare origin for GRB 200415A}
\shortauthors{Zhang et al. 2020}

\begin{document}

\title{Magnetar giant flare origin for GRB 200415A inferred from a new scaling relation and its implications}
\author[0000-0001-6863-5369]{Hai-Ming Zhang}
\affil{School of Astronomy and Space Science, Nanjing University, Nanjing 210023, China; xywang@nju.edu.cn; ryliu@nju.edu.cn}
\affil{Key laboratory of Modern Astronomy and Astrophysics (Nanjing University), Ministry of Education, Nanjing 210023, China}
\author[0000-0003-1576-0961]{Ruo-Yu Liu}
\affil{School of Astronomy and Space Science, Nanjing University, Nanjing 210023, China; xywang@nju.edu.cn; ryliu@nju.edu.cn}
\affil{Key laboratory of Modern Astronomy and Astrophysics (Nanjing University), Ministry of Education, Nanjing 210023, China}
\author[0000-0002-1766-6947]{Shu-Qing Zhong}
\affil{School of Astronomy and Space Science, Nanjing University, Nanjing 210023, China; xywang@nju.edu.cn; ryliu@nju.edu.cn}
\affil{Key laboratory of Modern Astronomy and Astrophysics (Nanjing University), Ministry of Education, Nanjing 210023, China}
\author[0000-0002-5881-335X]{Xiang-Yu Wang}
\affil{School of Astronomy and Space Science, Nanjing University, Nanjing 210023, China; xywang@nju.edu.cn; ryliu@nju.edu.cn}
\affil{Key laboratory of Modern Astronomy and Astrophysics (Nanjing University), Ministry of Education, Nanjing 210023, China}

\begin{abstract}

Soft gamma-ray repeaters (SGRs) are mainly a Galactic population and originate from neutron stars with intense ($B\simeq 10^{15}{\rm \ G}$) magnetic fields ('magnetars'). Occasionally, a giant flare  occurs with enormous intensity, displaying a short hard spike, followed by a weaker, oscillatory phase which exhibits the rotational period of the neutron star. If the magnetar giant flares occur in nearby galaxies, they would appear as cosmic short-hard gamma-ray bursts (GRBs) without detecting the weak oscillatory phase. Recently, a short-hard GRB named GRB 200415A was detected, with a position coincident with the Sculptor Galaxy (NGC 253), rasing the question whether it is a classic short GRB or a magentar giant flare. Here we show that magnetar giant flares follow a scaling relation between the spectral peak energy and the isotropic energy in  $1\,{\rm keV} - 10\,{\rm MeV}$, i.e., $E_{\rm p}\propto E_{\rm iso}^{1/4}$, and locate in a distinct region of  the $E_{\rm p}-E_{\rm iso}$  plane from that of classic short GRBs. The relation can be well understood in the  model that giant flares arise from the photosphere emissions of  relativistically expanding fireball. GRB 200415A, together with two other candidate giant flares (GRB 051103 and GRB 070201) follow this relation, which strongly favor the giant flare origin of these GRBs.  The GeV emission detected by \emph{Fermi}/LAT from GRB 200415A at $18-285 \,$s can also be explained in the giant flare scenario. The total energy in the GeV emission implies a baryon load of $\sim 10^{23}{\rm g}$ in the giant flare fireball of GRB 200415A.

\end{abstract}


%%%%%%%%%%%%%%%%%%%%%%%%%%%%%%%%%%%%%%%%%%%%%%%%%%%%%%%%%%%%%%%%
\section{Introduction}           %% first-level sections will be auto-capitalized
\label{sect:intro}

Giant flares (GFs), with total energies  in excess of $10^{44}{\rm erg}$, have been detected from three known SGRs. GRB 790305 is the first GF  detected from SGR 0526-66 \citep{Mazets1979a,Cline1980}. After $\sim20$ years later, two  more GFs  were detected, GRB 980827 from SGR 1900+14 \citep{Hurley1999a,Feroci1999,Mazets1999b} and GRB 041227 from SGR 1806-20 \citep{Hurley2005,Frederiks2007a}.  An intermediate flare (GRB 980618), with an energy of $\sim 10^{43}{\rm \ erg}$ was detected from SGR 1627-41 \citep{Mazets1999a,Hurley1999b}.
These events are now considered to be a very rare {type of} astrophysical phenomenon and completely different from  classic short gamma-ray bursts (GRBs), which result from mergers of compact objects and occur at cosmological distance typically.
On the other hand, however, if a magnetar giant flare occurs in an external galaxy, only the initial peak of a GF would be detectable, and thus the GF would resemble a several hundred millisecond long, hard-spectrum GRB \citep{Hurley2005,Palmer2005}. Indeed, two candidate giant flares have been suggested, i.e.,  GRB 051103, spatially coincident with the galaxy M81 \citep{Golenetskii2005}, and GRB 070201, coincident with the galaxy M31 \citep{Mazets2008}.  The chance coincidence probability between the IPN localization of GRB 051103 (and GRB 070201) and a nearby galaxy is low ($\sim 1\%$; \cite{Svinkin2015}).

At 08:48:05.56 UT on 15 April 2020, the Fermi Gamma-Ray Burst Monitor (GBM)
triggered and located GRB 200415A,
which was also detected by \emph{Fermi}/LAT \citep{Bissaldi2020,Fermi2020,Omodei2020}.
Interestingly, the burst is localized by the IPN and the localization (a 274 arcmin$^2$ error box) shows that the Sculptor Galaxy (NGC 253) is inside the IPN box \citep{Svinkin2020}.  Thus, it is possible that GRB 200415A is a GF from the Sculptor Galaxy \citep{Svinkin2020,Frederiks2020,Pozanenko2020,Yang2020}. However, without identifying the oscillatory phase or SGR activity, these candidates can not be confirmed reliably to be  GFs.
In this paper, we study whether the GRB 200415A is a GF  and study the origin of the GeV emission.

\section{GBM Data Reduction and Analysis}
\begin{figure*}
\begin{center}
\includegraphics[angle=0,scale=0.35]{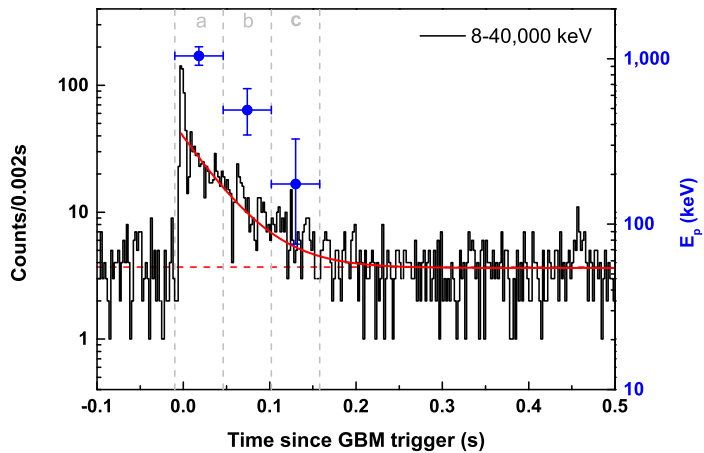}
\caption{Light curve and $E_{p}$ evolution of GRB 200415A in the GBM energy range (8--40,000 keV) on semilogarithmic scale. The red dashed line indicate the background counts. The burst decays quasi-exponentially with an e-folding time $\tau=42.5$ ms (the red line). The gray dashed lines indicate the three intervals where the spectral peaks were measured. }
\label{lc2}
\end{center}
\end{figure*}
%The GBM data of this GRB is downloaded from the public science support center at the official Fermi Web site\footnote{\url{https://fermi.gsfc.nasa.gov/ssc/data/}}.
{GRB 200415A triggered the Fermi/GBM at 08:48:05.56 UT on 15 April 2020 ($T_0$) \citep{Fermi2020}.}
We select the trigger detectors, e.g., NaI detectors n0 \& n1 and BGO detector b0.
We first extract the time integrated spectrum in the peak region (i.e., from $T_0-0.010$ to $T_0+0.158$ s).
We used the software package XSPEC (version 12.10) to perform spectral fitting. We find that the spectrum can be successfully fitted by a power law function with an exponential high-energy cutoff (hereafter, cutoff power law or CPL model), i.e., $N(E)=AE^{-\alpha}{\rm exp}[-(2-\alpha)E/E_p]$, {where the $N(E)$ represents the photon number spectrum}. The goodness of statistics for the fit is {$PGSTAT= 327.83$ and the degree of freedom is $dof =359$}. The power law index is {$-0.05_{-0.10}^{+0.10}$ and the spectral peak energy, $E_{\rm p}$, is $916.3_{-112.8}^{+124.5} \rm \ keV$.} The average flux in this time interval is {$(5.08\pm0.56) \times 10^{-5} \ {\rm erg \ cm^{-2} \ s^{-1}}$ {(between $1 \rm \ keV$ and $10 \rm \ MeV$) and the total fluence is $(8.53\pm0.94) \times 10^{-6} \ {\rm erg \ cm^{-2}}$}.
Assuming that the source of GRB 200415A is situated in NGC 253 at a distance of 3.5 Mpc, the measured value of the fluence corresponds to a
total isotropic energy is $E_{\rm iso}=(1.25\pm0.13) \times 10^{46} \ {\rm erg}$.
We also tested the the Band function model in the spectral analysis \citep{Band1993}. We found that it also can successfully fit the spectrum, but no statistically significant high-energy power-law tail is found ($\beta<-2.95$ is not constrained well).
%To check whether the Band function fit is overfitting (since it has one extra parameter), we employ a Bayesian information criterion (BIC) to check its statistical confidence. The comparison of the two models leads to $\triangle BIC =5.66$ (Band function model has the higher BIC). As suggested by \citet{lv2017}, such a $\triangle BIC$ value indicates  evidence against the higher BIC model, i.e., the Band function model.
Therefore, we favorably choose the cutoff power law model as the best-fit model in our analysis for this GRB.

The light curve of GRB 200415A observed in the GBM energy range (8--40,000 keV) with a time resolution of 2 ms is shown in Figure \ref{lc2} on semilogarithmic scale. The light curve has a  shape of a single short pulse {($t_{\rm burst}\sim168 \rm \ ms$)} with a steep leading edge {($t_{\rm rise}\leq6 \rm \ ms$)} and shows a quasi-exponential decay with an e-folding time of $\tau=42.5 \rm \ ms$.  As shown in Figure \ref{lc2}, we divided this time interval into three segments to study its spectral evolution.
We found that the spectrum displayed a strong hard-to-soft spectral evolution, which is consistent with the spectral characteristics of both GFs and classic GRBs \citep{Hurley2005,Preece1998}.

\section{The $E_{\rm p}$-$E_{\rm iso}$ Correlation for magnetar giant flares}

\subsection{The GF sample }
\begin{table*}
\caption{\label{tab:GFs} Properties of magnetar GFs and Candidates}
%\setlength{\tabcolsep}{2pt}

%\begin{tiny}
\begin{center}
\begin{tabular}{clcccc}
\hline

GRB & SGR or associated galaxy &  $D_{L}$  & $E_{\rm iso}$ & $E_{\rm p}$ \\
 &  &  kpc  & $\rm erg$ & $\rm keV$ \\
\hline
790305\tablenotemark{a}&  SGR 0526--66 (LMC) & $49.97\pm0.19$  &$(6.99\pm0.05)\times10^{44}$ & $520.0\pm100.0$\\ 		
980618\tablenotemark{b}&  SGR 1627--41       & $11.0\pm0.3$  &$(1.08\pm0.06)\times10^{43}$ & $125.0\pm25.0$\\ 	
980827\tablenotemark{c}&  SGR 1900+14       & $12.5\pm1.7$  &$(2.99\pm0.81)\times10^{44}$ & $240.0\pm20.0$\\ 	
041227\tablenotemark{d}&  SGR 1806--20       & $8.7_{-1.5}^{+1.8}$  &$(7.23_{-2.49}^{+2.99})\times10^{45}$ & $480.0\pm40.0$\\
051103\tablenotemark{e}&  M81 & $3630\pm340$  &$(7.12\pm1.33)\times10^{46}$ & $\sim900.0$\\ 	
070201\tablenotemark{f}&  M31 & $744\pm33$  &$(1.36\pm0.12)\times10^{45}$ & $296.0_{-32.0}^{+38.0}$\\
200415A\tablenotemark{g}& Sculptor Galaxy (NGC 253)  & $3500\pm200$  &$(1.25\pm0.14)\times10^{46}$ & $916.3_{-112.8}^{+124.5}$\\	
\hline
\end{tabular}
\end{center}
\tablenotetext{a}{\cite{Pietrzy2013,Mazets1982,Mazets2008} }
\tablenotetext{b}{\cite{Corbel1999,Mazets1999a} }
\tablenotetext{c}{\cite{Davies2009,Hurley1999a,Tanaka2007}}
\tablenotetext{d}{\cite{Bibby2008,Palmer2005}}
\tablenotetext{e}{\cite{Freedman1994,Frederiks2007b,Mazets2008}}
\tablenotetext{f}{\cite{Vilardell2010,Mazets2008}}
\tablenotetext{g}{\cite{Rekola2005} and this work.}

%\end{tiny}
\end{table*}

GRB 790305 is a  GF detected from SGR 0526-66 \citep{Mazets1979a}. Its spectrum can be fitted by an optically thin thermal bremsstrahlung (OTTB) model $ dN(E)/dE\propto E^{-1}\exp(-E/kT)$, with $kT=520\pm100 \rm \ keV$ \citep{Mazets1981,Mazets1982,Fenimore1996}). For this spectrum, the peak of the  $E^2 dN(E)/dE$ spectrum is at $E_{\rm p}=kT$.
GRB 980827 is a GF detected from SGR 1900+14 \citep{Hurley1999a}. Its spectrum is also well represented by an OTTB model, so we take the peak energy $E_{\rm p}$ (averaged over a 1-s interval) as $kT=240\pm20 \rm \ keV$ (the error bar is obtained from Figure 1 (panel b) of \cite{Hurley1999a}). GRB 041227 is a GF detected from SGR 1806-20 \citep{Hurley2005,Palmer2005,Frederiks2007a}. The spike's intensity drove all X- and $\gamma$-ray
detectors into saturation, but particle detectors aboard RHESSI and Wind made reliable measurements \citep{Hurley2005}. The SOPA and ESP instruments located on
geosynchronous satellites also measured the flux and spectrum\citep{Palmer2005}. {The SOPA data of this GF  were fitted with a CPL model, and we take the spectral peak energy and flux   from \cite{Palmer2005}.} GRB 980618 is an intermediate flare detected from SGR 1627-41 and its spectrum can be modelled by an OTTB spectrum \citep{Mazets1999a}. From Figure 3 of \citet{Mazets1999a}, we can infer that the spectral peak of the $E^2 dN(E)/dE$ spectrum is equal to $kT=100-150 \rm \ keV$ (here we set $E_{\rm p}$ to $125\pm25 \rm \ keV$). The hard spectrum of this flare, together with  a total energy of $10^{43}{\rm \ erg}$, makes it quite similar to giant flares. Indeed,  \citet{Mazets2008} consider this flare as a giant flare.  For these reasons, we include the flare from SGR 1627-41 in the correlation analysis.
Two GF candidates, GRB 051103 \citep{Frederiks2007b} and 070201 \citep{Mazets2008} are both well fitted by the CPL model. The spectral peak energies of these two GFs are taken  from \cite{Mazets2008}.

The spectral peak energy and the total isotropic energy of GRB 200415A, together with those of four GFs and two GF candidates, are summarized in Table \ref{tab:GFs}. {Note that, in Table \ref{tab:GFs}, we also give the uncertainty in the total isotropic energy of GFs due to the uncertainty in the source distance.}

\subsection{$E_{\rm p}$-$E_{\rm iso}$ Correlation}
\begin{figure*}
\begin{center}
\includegraphics[angle=0,scale=0.4]{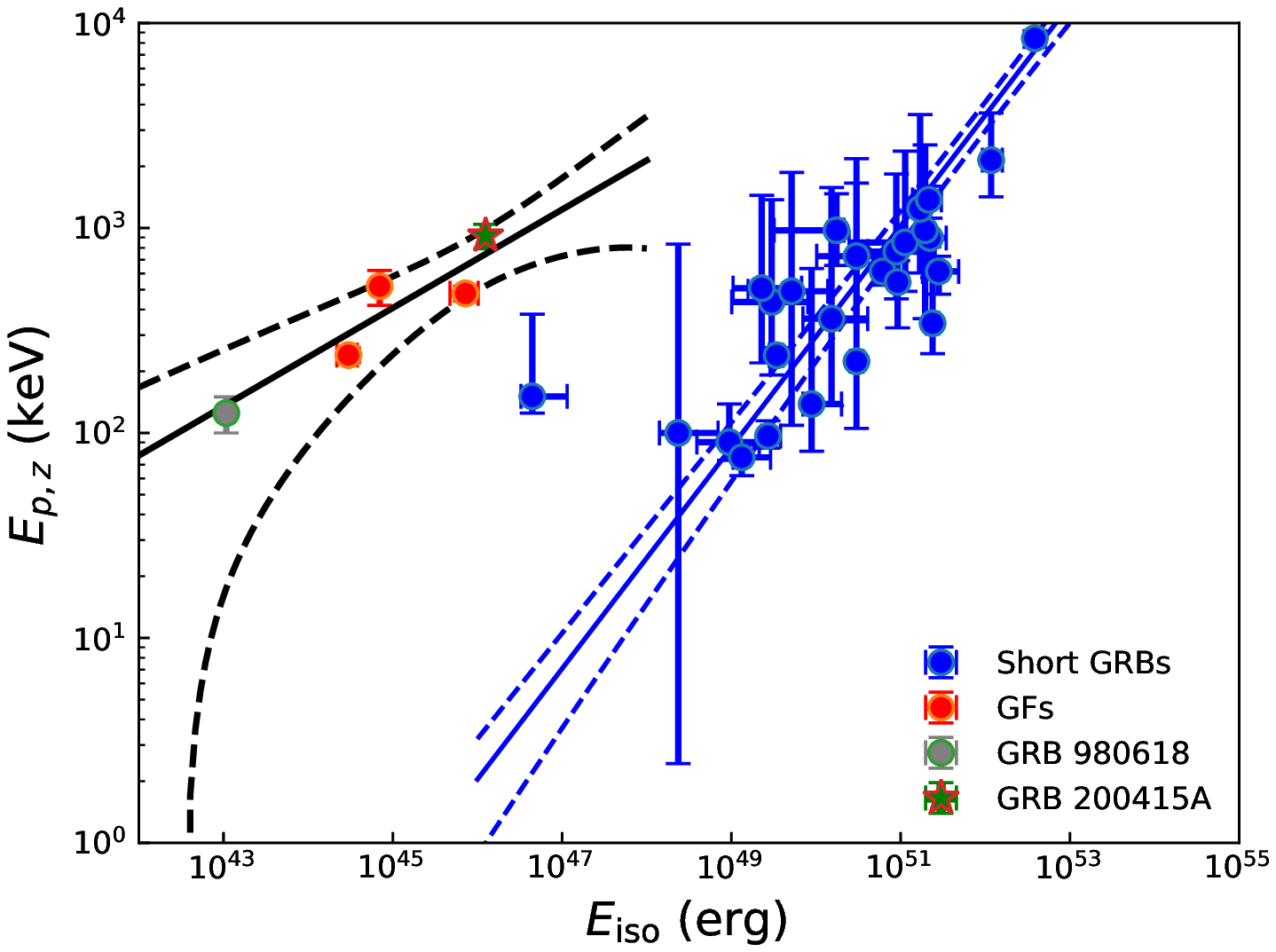}
\includegraphics[angle=0,scale=0.4]{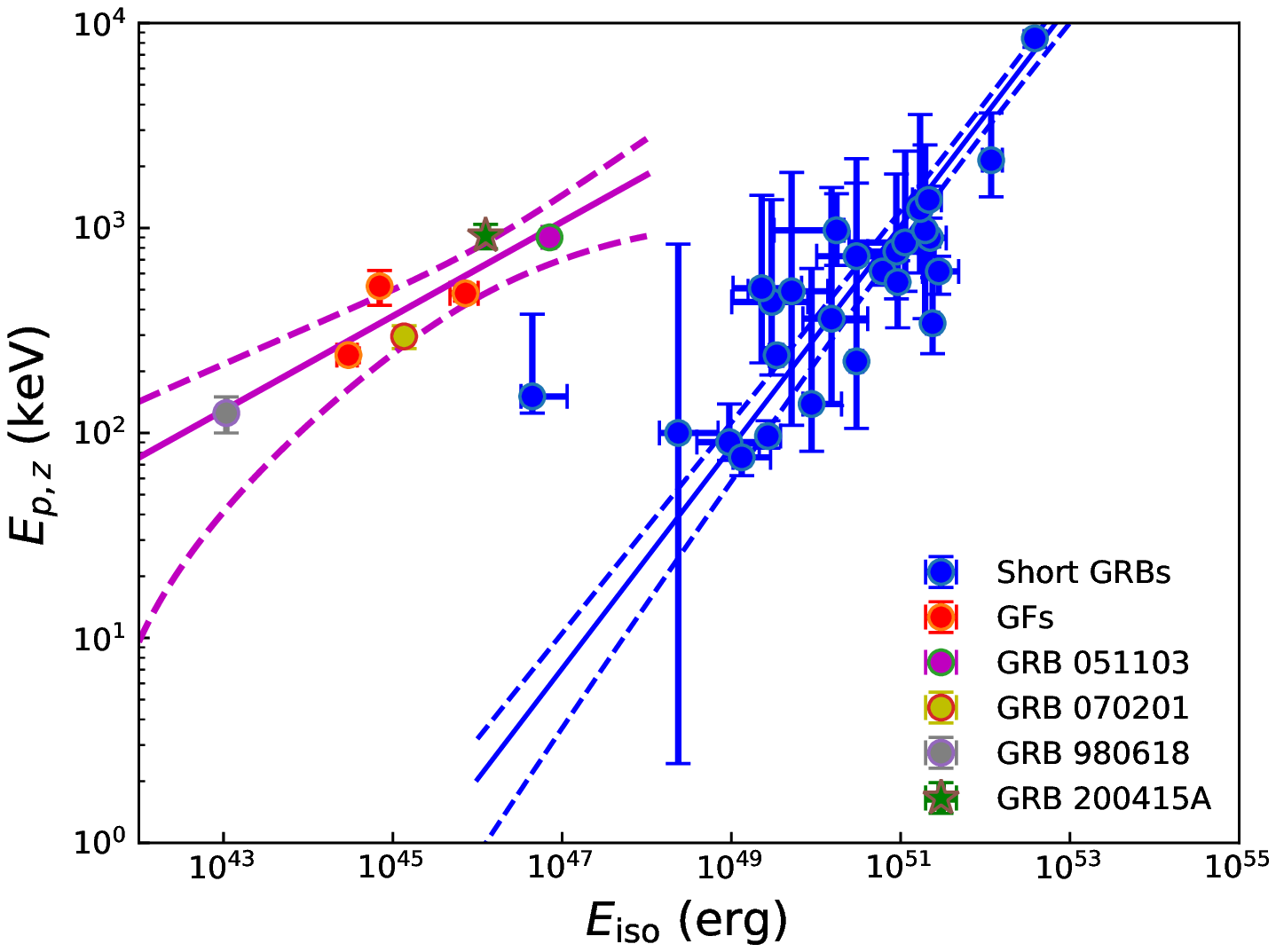}
\caption{GRB 200415A in the $E_{\rm {p,z}}$-$E_{\rm iso}$ correlation diagram. The blue data points represent short GRBs, which are taken from \citet{Zhang2018}. The GFs (red), intermediate flare GRB 980618 from SGR 1627-41 (green) and GF candidates (yellow and magenta) marked as the color data points are derived from the references as shown in Table \ref{tab:GFs}.
The green star represents  GRB 200415A.
The blue solid line is the best fit correlation: $\log (E_{\rm {p,z}}/{\rm keV})=(3.55\pm0.07)+(0.54\pm0.04)\log(E_{\mathrm{iso}}/10^{52}\rm erg)$ for short GRBs.
The blue dashed lines represent the $1\sigma$ region of the correlation.
Left panel:
The black solid line represents the best-fit correlation: $\log (E_{ \rm {p}}/{\rm keV})=(2.85_{-0.05}^{+0.04})+(0.24_{-0.03}^{+0.04})\log (E_{\mathrm{iso}}/10^{46}\rm erg)$ for considering three identified GFs, the intermediate flare and GRB 200415A.
The black dashed lines represent $3\sigma$ region of the correlation. Right panel: The magenta solid line is the best-fit correlation for considering three identified GFs, the intermediate flare and three candidates (including GRB 051103, 070201 and 200415A):
$\log (E_{ \rm {p}}/{\rm keV})=(2.80_{-0.03}^{+0.02})+(0.23_{-0.02}^{+0.03})\log (E_{\mathrm{iso}}/10^{46}\rm erg)$.
The magenta dashed lines represent $3\sigma$ regions of the correlation.
{The discrepant blue data point of short GRB at $E_{\rm iso}\sim10^{46.5} \rm \ erg$ represents  GRB 170817A/GW 170817.}
}
\label{correlation}
\end{center}
\end{figure*}

With the observed properties of these magnetar GFs and candidates, we study relation between the spectral peak energy $E_{\rm p,z}$ and the isotropic energy $E_{\rm iso}$ of them. {Since the samples in this work is small, the scaling relations are estimated by ordinary least squares for an ensemble of datasets where the data point are selected by bootstrap resampling and have positions jiggled by values consistent with their uncertainties (considering both the uncertainties of $E_{\rm p}$ and $E_{\rm iso}$). The best-fit results of the slopes and intercepts are set to be the median of the distribution of the ensemble, and the uncertainties of them are calculated by $1\sigma$ confidence intervals, which is obtained from the distribution of the ensemble.

We first use the three {identified} GFs and the intermediate flare for the study.
The best-fit relation between the peak energy $E_{\rm p}$ and the total isotropic energy $E_{\rm iso}$  is $\log (E_{ \rm {p}}/{\rm keV})=(2.78_{-0.05}^{+0.06})+(0.22_{-0.05}^{+0.04})\log (E_{\mathrm{iso}}/10^{46}\rm erg)$, with a Pearson correlation coefficient of $\kappa=0.91$  and a chance probability of $p=0.09$.
If we  consider only the three identified GFs, the correlation becomes weaker, but the slope is  consistent with the above analysis, i.e.,  $\log (E_{ \rm {p}}/{\rm keV})=(2.73_{-0.05}^{+0.06})+(0.15_{-0.05}^{+0.06})\log(E_{\mathrm{iso}}/10^{46}\rm erg)$, with a Pearson correlation coefficient of $\kappa=0.64$. }
As shown in Figure \ref{correlation}, plotting GRB 200415A onto the $E_{ \rm p}$ vs. $E_{\rm iso}$ plane, we find that GRB 200415A is in good agreement with this relation and far away from the short GRB population. This indicates that GRB 200415A is more likely to be a magnetar GF from the Sculptor Galaxy.

Including the three GFs, the intermediate flare, and GRB 200415A to do the re-fitting, we find that the best-fit relation gives a tighter relation: $\log (E_{ \rm {p}}/{\rm keV})=(2.85_{-0.05}^{+0.04})+(0.24_{-0.03}^{+0.04})\log (E_{\mathrm{iso}}/10^{46}\rm erg)$, with a Pearson correlation coefficient of $\kappa=0.93$  and a chance probability of $p=0.02$.
Considering also the other two candidate GFs (GRB 051103 and GRB 070201) lying within the $3\sigma$ of the track of the GF population, we include these sources to do the re-fitting. We find that the best-fit relation gives a tighter relation: $\log (E_{ \rm {p}}/{\rm keV})=(2.80_{-0.03}^{+0.02})+(0.23_{-0.02}^{+0.03})\log (E_{\mathrm{iso}}/10^{46}\rm erg)$, with a Pearson correlation coefficient of $\kappa=0.93$  and a chance probability of $p=0.003$. This correlation suggests that these sources and the identified GFs are the same class of sources.
This relation is quite different from the correlation for short GRBs{\footnote{If GRB 200415A is not physically associated with the Sculptor galaxy, but is instead a chance
coincidence and originates from an unrelated galaxy at a cosmological distance, then it could satisfy the $E_{p,z}-E_{iso}$ relation for a redshift of $z>0.025$.}}, i.e., the "Amati relation" \citep{Amati2002} with $\log (E_{\rm {p,z}}/{\rm keV})=(3.55\pm0.07)+(0.54\pm0.04)\log(E_{\mathrm{iso}}/10^{52}\rm erg)$ \citep{Zhang2018}, where $E_{\rm {p,z}}=E_{\rm p}\rm (1+z)$.
As shown in Figure \ref{correlation},  GFs do not fall on the the Amati relation of short GRBs.
%As shown in Figure \ref{correlation},  GFs lie outside of the $3\sigma$ region of the Amati relation for short GRBs.
We also find that the two correlations do not agree at $>5\sigma$ level, indicating that GFs are distinct from short GRBs.
}

\subsection{Interpretation of the scaling relation}
Next we show that the  correlation between $E_{\rm p}$ and  $E_{\rm iso}$ for magnetar GFs can be understood in the framework of the fireball photosphere emission model. During a GF, strong shearing of a neutron star's magnetic field,
combined with growing thermal pressure, appears to have forced an
opening of the field outward \citep{Thompson2001}. {A huge amount of magnetic energy $E=10^{46}{\rm \ erg}$ is subsequently released at the surface of the neutron star with a radius of $R_0$ over a short period of $t=0.1{\rm \ s}$, leading to the formation of a hot fireball \citep{Thompson1995,Thompson2001,Nakar2005}, which is quite similar to the case of classic GRBs \citep{Meszros2000}.} The optical depth for pair production is extremely high, {regardless of the suppression of} the effective cross-section due to the large magnetic field of the magnetar \citep{Herold1979}. With such a large optical depth, a radiation-pairs plasma is formed at a thermal equilibrium with an initial temperature
\begin{equation}
kT_0=\left (\frac{E}{4\pi R_0^2 \sigma t}\right)^{1/4}=200 {\rm \ keV} E_{46}^{1/4} R_0^{-1/2} t_{-1}^{-1/4},
\label{eq:T_0}
\end{equation}
where $\sigma$ is the Stephan-Boltzmann constant. This plasma expands under its own pressure, {launching a relativistic outflow}. As the optically thick (adiabatic) outflow expands, the comoving internal energy drops as $e'\propto V'^{-4/3}\propto n'^{4/3}$, where $V'$ is the comoving volume and $n'$ is the comoving baryon number density. The baryon bulk Lorentz factor increases as $\Gamma\propto R$ and the comoving temperature drops $T'\propto R^{-1}$. The $e^{\pm}$ pairs drop out of equilibrium {\citep{Paczynski1986}} at $T'_p=17 {\rm \ KeV}$ at a radius of $R_p=R_0 (T_0/T'_p)$.
This is the radius of an $e^{\pm}$ pair photosphere. If the outflow carries enough baryons,  the baryonic electrons lead to a photosphere larger than $R_p$. As long as the outflow remains optically thick, it is radiation dominated and continues to expand with $\Gamma\propto R$. For large baryon loads,  $\Gamma$ reaches the saturation  value $\eta$ at a  radius $R_s=\eta R_0$, where the baryon load is parameterized by $\eta=E/M c^2$. Above the saturation radius, the flow continues to coast with $\Gamma=\eta$.

On the other hand, for low baryon loads, a baryonic electron photosphere appears in the accelerating portion $\Gamma\propto R$.  An electron scattering photosphere is defined by $\tau=\sigma_{\rm T} n'  R_{ph}/\Gamma=1$, where $\tau$ is the Thomson optical depth, $\sigma_{\rm T}$ is the Thomson cross-section, $n'=(L/4\pi m_p c^3 \Gamma \eta)$ is the comoving baryon density, $r/\Gamma$ is a typical comoving length, $m_p$ is the proton mass, and $L$ is the total power of the burst.  There is a critical value of $\eta$ at which $R_{ph}=R_s$, i.e.,
\begin{equation}
\eta_*=\left( \frac{L\sigma_{\rm T} }{4\pi m_p c^3 R_0}\right)^{1/4}=100 L_{47}^{1/4} R_{0,6}^{-1/4} .
\label{eq:eta}
\end{equation}
Thus, for low baryon loads where $\eta\ge \eta_*$, the outflow becomes optically thin in the accelerating portion.
The outflow emits a quasi-blackbody spectrum as it became optically thin, with a spectral temperature comparable to the temperature at its base, because declining temperature in the outflow is compensated by the relativistic blueshift, i.e., $T_{ph}=\Gamma T'_{ph}=T_0$.
The observed photospheric thermal luminosity is $L_{ph}=4\pi R^2 \Gamma^2 \sigma T'^4_{ph}=4\pi R_0^2 \sigma T_{ph}^4$.
Thus, the observer-frame photospheric temperature is
\begin{equation}
kT_{\rm ph}=\left (\frac{E_{\rm iso}}{4\pi R_0^2 \sigma t }\right)^{1/4}=200 {\rm \ KeV} E_{\rm iso,46}^{1/4} R_0^{-1/2} t_{-1}^{-1/4},
\label{eq:T_ph}
\end{equation}
where $E_{\rm iso}=L_{\rm ph} t$ and we have assumed a duration of $t=0.1{\rm \ s}$ for the all the GFs. The peak energy of the spectrum is at $E_{\rm p}=2.8 \ kT_{\rm ph}$. It is remarkable to see that Eq.(\ref{eq:T_ph}) agrees well with the
$E_{\rm p}-E_{\rm iso}$ relation that we found for GFs.

The spectra of GRB 200415A and some other GFs are not purely blackbody and better modelled by CPL spectrum. This can be interpreted as {being due to} the extra contribution by some other radiation components to the low-energy spectrum, such as multi-temperature blackbody emission or synchrotron emission arising from some shocks in the relativistic outflow \citep{Meszros2000}. There is evidence for multi sub-pulses and fast variability in the initial spike in this and other GFs. Different sub-pulses have different temperatures, so  multi-temperature blackbody is naturally expected. The fast variability implies unsteady outflow, so internal shocks could occur, which may produce a flat synchrotron spectrum. If this synchrotron spectrum is subdominant compared with the photosphere emission,  the low-energy spectrum will be flatter while the peak energy remains unchanged \citep{Meszros2000}. A detailed study of the spectrum considering these effects is beyond the scope of the present paper.

Note that Eq.(\ref{eq:T_ph}) holds only when $R_{ph}<R_s$, corresponding to $\eta>\eta_*$. If $R_{ph}>R_s$ ($\eta<\eta_*$), the observer-frame photospheric temperature and photospheric thermal luminosity  evolve as $T_{ph}\propto R^{-2/3}$ and $L_{ph}\propto R^{-2/3}$, respectively. In this case, the relation is different from Eq.(\ref{eq:T_ph}). That is, the temperature (or the peak energy of the spectrum) would be smaller than observed if the flow is baryon-rich, and therefore, in the present case, the outflow must be baryon-poor. This implies that baryon load in the outflow of GRB 200415A should satisfy the condition $\eta>\eta_*$.

The outflow reaches a Lorentz factor of $\Gamma=R_{ph}/R_0=\eta_*(\eta/\eta_*)^{-1/3}$ at the photosphere. Beyond the photosphere, most of the electrons above the photosphere can still scatter with a decreasing fraction of free-streaming photons  as long as the comoving Compton drag time is less than the comoving expansion time \citep{Meszros2000}. As a result, for $\eta>\eta_*$, the terminal Lorentz factor of the outflow is $\eta_*$ (instead of $\eta$).  The energy that remains in the baryons of the outflow is $E_b=E (\eta_*/\eta)$. The energy in the afterglow emission of GFs could place a limit on the energy in the final baryonic ejecta{\footnote{The kinetic energy of the outflow in the giant flare of SGR 1806-20 is constrained by the radio afterglow \citep{Gaensler2005,Gelfand2005,Granot2006}.}}. The fluence of the GeV emission of GRB 200415A is about $(2.9\pm0.2) \times10^{-6}{\rm \ erg \ cm^{2}}$ in the time interval $0-300$\,s, which is only a factor of 3 smaller than the fluence of the GF. This implies that $\eta$ should not be much larger than $\eta_*$. As a result, we estimate that the baryon load in the GF of GRB 200415A is about $M\sim E/(\eta_* c^2)\sim 10^{23}{\rm g}$.

\section{Origin of the GeV emission}
\begin{figure*}
\begin{center}
\includegraphics[angle=0,scale=0.5]{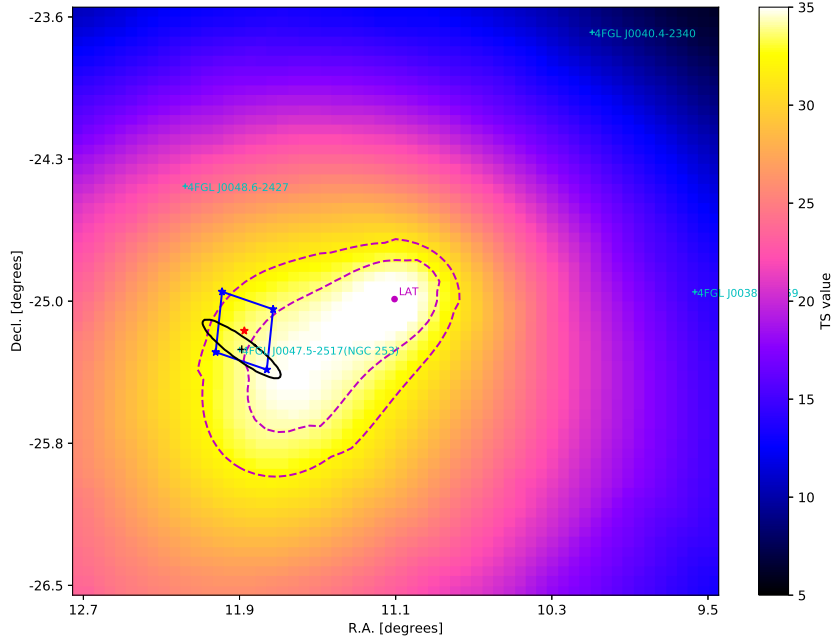}
\caption{$3\degr \times 3\degr$ TS map of the GRB 200415A in the energy band 0.1-10 GeV. The green crosses represent the positions of the 4FGL point sources.
Black contour depicts the optical emission from the whole NGC 253 with contour level of constant surface brightness of 25 mag $\rm arcsec^{-2}$ as used in \citet{Pence1980}, the black cross represents the the optical centre of the NGC 253.
The magenta point indicates the best localization of GRB 200415A. The two magenta dashed circles represent the localization contours of GRB 200415A in 68\% and 90\% confidence levels. The red (center) and blue stars (corners) represent the IPN error box of GRB 200415A \citep{Svinkin2020}.
This map has been created for a pixel size of 0.05, smoothed by gaussian kernel ($\sigma=0.35\degr$). The color bar represents the value of TS per pixel.
}
\label{tsmap}
\end{center}
\end{figure*}
\begin{figure*}
\begin{center}
\includegraphics[angle=0,scale=0.4]{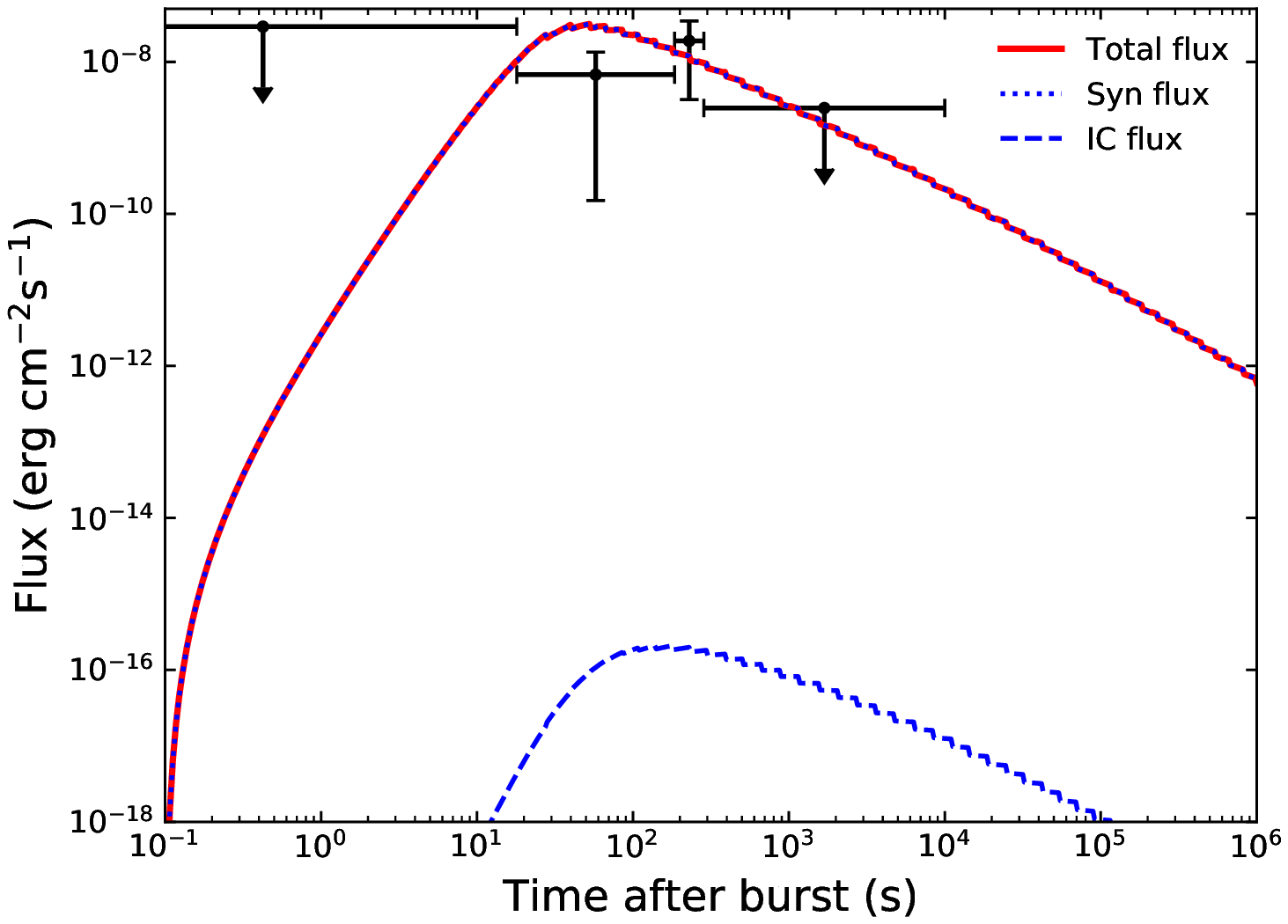}
\includegraphics[angle=0,scale=0.4]{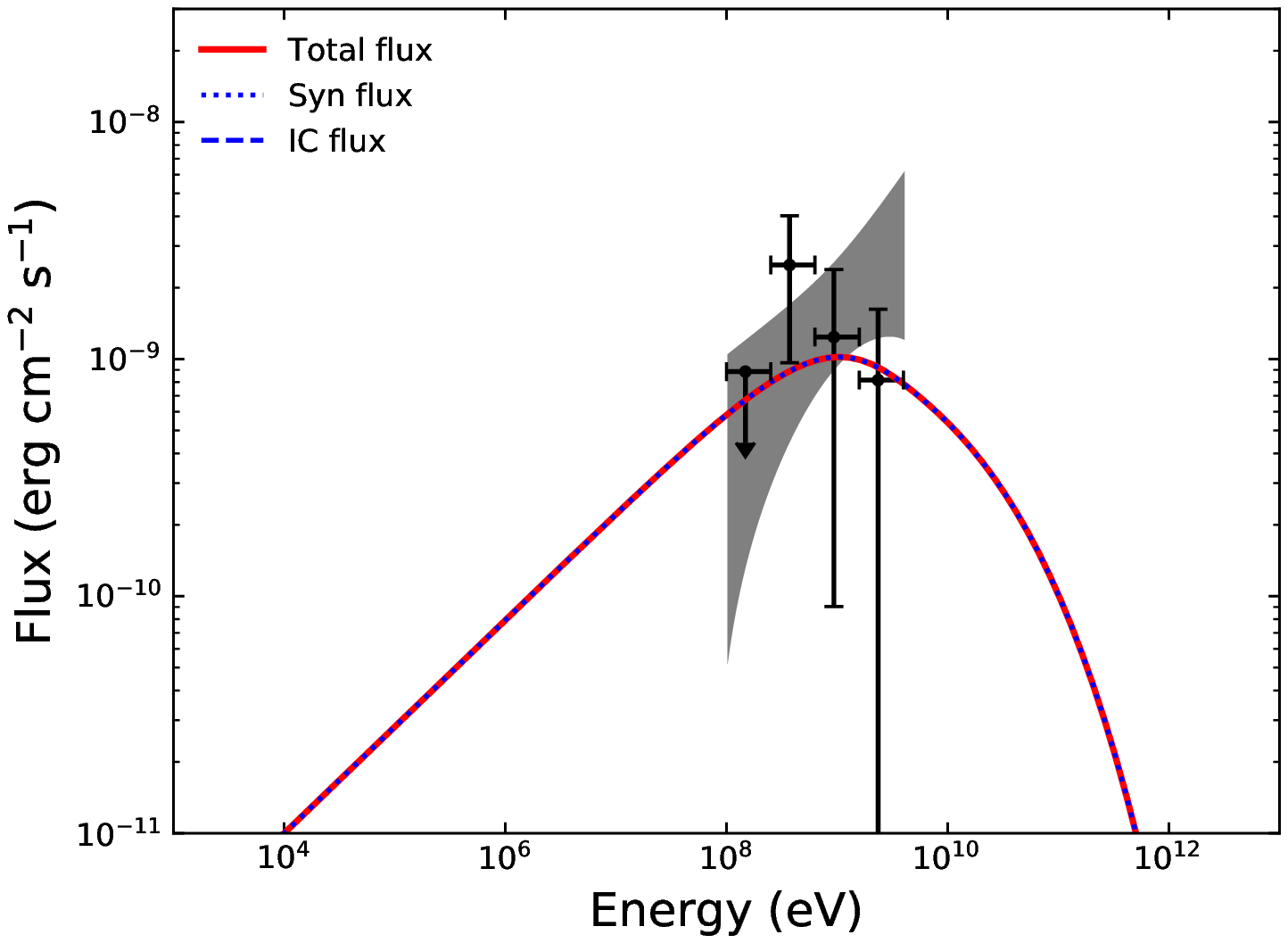}
\caption{Left panel: modeling of the  light curve of the GeV emission of GRB 200415A. The dotted curve and dashed curve represent the synchrotron component and SSC component, respectively. Right panel: modeling of the  spectrum of the GeV emission of GRB 200415A at t =18--285 s. The black data points represent the GeV flux spectra of GRB 200415A. The grey butterfly shows the best fit power-law model with $1\sigma$ error.
The parameters used in the fitting are $E = 1.25\times10^{46}\rm \ erg$, $n = 1\times10^{-5}\rm \ cm^{-3}$, $\epsilon_{e} = 0.95$, $\epsilon_{B} = 0.08$, $\Gamma_{0} = 100$, and $p = 2.1$.}\label{sed}
\end{center}
\end{figure*}
Remarkably, this event was detected by \emph{Fermi}/LAT in the GeV band \citep{Omodei2020}.
The \emph{Fermi}/LAT data for the GRB 200415A was taken from the Fermi Science Support Center\footnote{\url{https://fermi.gsfc.nasa.gov}}. The data analysis was performed using the publicly available software \textit{fermitools} (ver. 1.0.0) with the unbinned likelihood analysis method. Assuming a power-law spectrum of the burst, we estimated the best-fit \emph{Fermi}/LAT position of GRB 200415A with the tool \emph{gtfindsrc}: (11.10$\degr$, -25.03$\degr$) with circular errors of 0.41$\degr$  and 0.68 $\degr$ (statistical only), respectively at 68\% and 90\% confidence levels.
In order to reduce the contamination from the $\gamma$-ray Earth limb, the maximum zenith angle is set to be 90$\degr$.
The \emph{P8R3$\_$TRANSIENT020$\_$V2} set of instrument response functions is used. %for transient sources
Since the highest-energy photon of GRB 200415A is a 1.72 GeV event, which is observed 284 seconds after the GBM trigger, we only consider the events with energies from 100 MeV to 10 GeV.
Taking into the consideration of background point-like sources, Galactic diffuse and isotropic emission, we found the test statistic (TS) of the burst is 35.9 ($\sim6 \sigma$) at $T_{0}$ to $T_{0}+300$ s. The TS value is defined as $\rm {TS}=2(\ln\mathcal{L}_{1}-\ln\mathcal{L}_{0})$, where $\mathcal{L}_{0}$ is the likelihood of background (null hypothesis) and $\mathcal{L}_{1}$ is the likelihood of the hypothesis for adding the burst. The averaged flux is $(9.52\pm0.61)\times 10^{-9}\ {\rm erg \ cm^{-2} \ s^{-1}}$ with a photon index $-1.46\pm0.37$.

As shown in Figure \ref{tsmap}, we compare the localization of LAT for GRB 200415A with the IPN error box and the position of NGC 253 (4FGL J0047.5--2517). We find that the NGC 253 is located inside the IPN error box, and the error box center is offset by 5.93 arcmin from the NGC 253. The IPN error box is almost inside the region of the localization contours of GRB 200415A at the 90\% confidence level. Considering also the temporal coincidence between the GeV emission and the giant flare, we suggest that the GeV emission of GRB 200415A originates from the NGC 253.

We perform modeling of the \emph{Fermi}/LAT data for GRB 200415A using the  numerical code \citep{Liu2013}. According to the standard afterglow model \citep{Sari1997}, the light curve for a given observed frequency ($\nu$) could be calculated as
$F(t,\nu)=F(t,\nu,E_{k},n,p,\varepsilon_{e},\varepsilon_{B},\Gamma_{0})$,
employing a numerical code developed in our previous work \citep{Liu2013}. $E_{k}$ is the isotropic kinetic energy of the GRB outflow, $n$ is the particle number density of the ambient medium, $p$ is the electron spectral index, $\varepsilon_{e}$ and $\varepsilon_{B}$ are the equipartition factors for the energy in electrons and magnetic field in the shock, and $\Gamma_0$ is the initial Lorentz factor of the outflow.

While GeV afterglow emission is commonly seen in short GRBs, it can not be taken as evidence against the magnetar GF origin. Indeed, GRB200415A is the first GF observed by \emph{Fermi}/LAT, which has started operation since 2008.
We suggest that the GeV emission may be produced by the shocks arising from the interaction between the relativistic outflow and the ambient medium. Generally there are two shocks, one is the forward shock expanding into the ambient medium and the other is the reverse shock expanding into the outflow ejecta. Both shocks could accelerate electrons,  producing synchrotron emission and inverse-Compton emission. Below we study the possibility of the forward shock  emitting the GeV afterglow emission.  Since the terminal Lorentz factor of the relativistic outflow after the acceleration is $\eta_*$, we take $\Gamma_0=100$ as a reference value for the initial Lorentz factor of the forward shock.
The modeling results of the afterglow light curve and SED for GRB 200415A are shown in Figure \ref{sed}.
The low density of the ambient medium is not surprising, since the pulsar wind and earlier SGR activity from the magnetar may have {created} a cavity around the pulsar. Indeed, it { has been} suggested that the Poynting flux emanating from the pulsar can excavate a bow-shock cavity around the pulsar with a size as large as fractions of a parsec \citep{Holcomb2014}. Interestingly, a bow-shock environment was discussed in \citet{Granot2006} to explain the radio nebula from the afterglow of the GF from SGR 1806-20.  The value of the electron equipartition factor $\epsilon_e=0.95$ seems to be  higher than that inferred for classic GRBs, but considering that the ambient medium around the pulsar may be enriched with pairs, such a value of $\epsilon_e$ may be reasonable for magnetar GFs \citep{Konigl2002}. We would like to point out that the model parameters cannot be reliably determined and the set of parameters we quoted is only an illustrative example.

\section{Conclusions}
Magnetar giant flares, if occurring in nearby galaxies, would appear as cosmic short-hard GRBs. Thus, identifying the nature of such kind of short GRBs becomes a challenge.
In this paper, we have shown that magnetar giant flares locate in a distinct region of the $E_{\rm p}-E_{\rm iso}$ plane and follow a scaling relation roughly as $E_{\rm p}\propto E_{\rm iso}^{1/4}$, quite different from those of classic short GRBs. Although the number  of giant flares in our sample is  small, the scaling relation is well consistent with the expectation of the standard model of magnetar giant flares, which ascribes the formation of the  hot fireballs as a consequence of  catastrophic instabilities
in  magnetars \citep{Thompson2001}. Such a scaling relation thus provides a powerful tool to distinguish between magnetar giant flares and classic short GRBs. Along this line, we suggest that GRB 200415A is a magnetar giant flare occurred in the nearby galaxy NGC 253.

\section*{Acknowledgments}
We thank the referee for the constructive report. The work is supported by the NSFC grants 11625312 and 11851304, and the National Key
R \& D program of China under the grant 2018YFA0404203.

%-----------------------------------------------------------------------------


\begin{thebibliography}{}

\bibitem[Amati et al.(2002)]{Amati2002} Amati, L., Frontera, F., Tavani, M., et al.\ 2002, \aap, 390, 81
\bibitem[Band et al.(1993)]{Band1993} Band, D., Matteson, J., Ford, L., et al.\ 1993, \apj, 413, 281
\bibitem[Bibby et al.(2008)]{Bibby2008} Bibby, J.~L., Crowther, P.~A., Furness, J.~P., et al.\ 2008, \mnras, 386, L23
\bibitem[Bissaldi et al.(2020)]{Bissaldi2020} Bissaldi, E., Briggs, M., Burns, E., et al.\ 2020, GRB Coordinates Network, Circular Service, No. 27587, 27587
\bibitem[Cline et al.(1980)]{Cline1980} Cline, T.~L., Desai, U.~D., Pizzichini, G., et al.\ 1980, \apjl, 237, L1
\bibitem[Corbel et al.(1999)]{Corbel1999} Corbel, S., Chapuis, C., Dame, T.~M., et al.\ 1999, \apjl, 526, L29
\bibitem[Davies et al.(2009)]{Davies2009} Davies, B., Figer, D.~F., Kudritzki, R.-P., et al.\ 2009, \apj, 707, 844
\bibitem[Duncan \& Thompson(1992)]{Duncan1992} Duncan, R.~C. \& Thompson, C.\ 1992, \apjl, 392, L9
\bibitem[Fenimore et al.(1996)]{Fenimore1996} Fenimore, E.~E., Klebesadel, R.~W., \& Laros, J.~G.\ 1996, \apj, 460, 964
\bibitem[Fermi GBM Team(2020)]{Fermi2020} Fermi GBM Team\ 2020, GRB Coordinates Network, Circular Service, No. 27579, 27579
\bibitem[Feroci et al.(1999)]{Feroci1999} Feroci, M., Frontera, F., Costa, E., et al.\ 1999, \apjl, 515, L9
\bibitem[Freedman et al.(1994)]{Freedman1994} Freedman, W.~L., Hughes, S.~M., Madore, B.~F., et al.\ 1994, \apj, 427, 628
\bibitem[Frederiks et al.(2007a)]{Frederiks2007a} Frederiks, D.~D., Golenetskii, S.~V., Palshin, V.~D., et al.\ 2007a, Astronomy Letters, 33, 1
\bibitem[Frederiks et al.(2007b)]{Frederiks2007b} Frederiks, D.~D., Palshin, V.~D., Aptekar, R.~L., et al.\ 2007b, Astronomy Letters, 33, 19
\bibitem[Frederiks et al.(2020)]{Frederiks2020} Frederiks, D., Golenetskii, S., Aptekar, R., et al.\ 2020, GRB Coordinates Network, Circular Service, No. 27596, 27596
\bibitem[Gaensler et al.(2005)]{Gaensler2005} Gaensler, B.~M., Kouveliotou, C., Gelfand, J.~D., et al.\ 2005, \nat, 434, 1104
\bibitem[Gelfand et al.(2005)]{Gelfand2005} Gelfand, J.~D., Lyubarsky, Y.~E., Eichler, D., et al.\ 2005, \apjl, 634, L89
\bibitem[Granot et al.(2006)]{Granot2006} Granot, J., Ramirez-Ruiz, E., Taylor, G.~B., et al.\ 2006, \apj, 638, 391
\bibitem[Golenetskii et al.(2005)]{Golenetskii2005} Golenetskii, S., Aptekar, R., Mazets, E., et al.\ 2005, GRB Coordinates Network, Circular Service, No. 4197, \#1 (2005), 4197
\bibitem[Herold(1979)]{Herold1979} Herold, H.\ 1979, \prd, 19, 2868
\bibitem[Holcomb et al.(2014)]{Holcomb2014} Holcomb, C., Ramirez-Ruiz, E., De Colle, F., et al.\ 2014, \apjl, 790, L3
\bibitem[Hurley et al.(1999a)]{Hurley1999a} Hurley, K., Cline, T., Mazets, E., et al.\ 1999a, \nat, 397, 41
\bibitem[Hurley et al.(1999b)]{Hurley1999b} Hurley, K., Kouveliotou, C., Woods, P., et al.\ 1999b, \apjl, 519, L143
\bibitem[Hurley et al.(2005)]{Hurley2005} Hurley, K., Boggs, S.~E., Smith, D.~M., et al.\ 2005, \nat, 434, 1098
\bibitem[Konigl \& Granot (2002)]{Konigl2002} K\"onigl, A. \& Granot, J., 2002, 574,134
\bibitem[Lipunov et al.(2020)]{Lipunov2020} Lipunov, V., Tyurina, N., Gorbovskoy, E., et al.\ 2020, GRB Coordinates Network, Circular Service, No. 27599, 27599
\bibitem[Liu et al.(2013)]{Liu2013} Liu, R.-Y., Wang, X.-Y., \& Wu, X.-F.\ 2013, \apjl, 773, L20
%\bibitem[L{\"u} et al.(2017)]{lv2017} L{\"u}, H.-J., L{\"u}, J., Zhong, S.-Q., et al.\ 2017, \apj, 849, 71
\bibitem[Mazets et al.(1979)]{Mazets1979a} Mazets, E.~P., Golentskii, S.~V., Ilinskii, V.~N., et al.\ 1979, \nat, 282, 587
\bibitem[Mazets \& Golenetskii(1981)]{Mazets1981} Mazets, E.~P. \& Golenetskii, S.~V.\ 1981, \apss, 75, 47
\bibitem[Mazets et al.(1982)]{Mazets1982} Mazets, E.~P., Golenetskii, S.~V., Gurian, I.~A., et al.\ 1982, \apss, 84, 173
\bibitem[Mazets et al.(1999a)]{Mazets1999a} Mazets, E.~P., Aptekar, R.~L., Butterworth, P.~S., et al.\ 1999a, \apjl, 519, L151
\bibitem[Mazets et al.(1999b)]{Mazets1999b} Mazets, E.~P., Cline, T.~L., Aptekar', R.~L., et al.\ 1999b, Astronomy Letters, 25, 635
\bibitem[Mazets et al.(2008)]{Mazets2008} Mazets, E.~P., Aptekar, R.~L., Cline, T.~L., et al.\ 2008, \apj, 680, 545
%\bibitem[M{\'e}sz{\'a}ros \& Rees(1997)]{Meszros1997} M{\'e}sz{\'a}ros, P. \& Rees, M.~J.\ 1997, \apj, 476, 232
\bibitem[M{\'e}sz{\'a}ros \& Rees(2000)]{Meszros2000} M{\'e}sz{\'a}ros, P. \& Rees, M.~J.\ 2000, \apj, 530, 292
\bibitem[Nakar et al.(2005)]{Nakar2005} Nakar, E., Piran, T., \& Sari, R.\ 2005, \apj, 635, 516
\bibitem[Omodei et al.(2020)]{Omodei2020} Omodei, N., Axelsson, M., Piron, F., et al.\ 2020, GRB Coordinates Network, Circular Service, No. 27586, 27586
\bibitem[Paczynski(1986)]{Paczynski1986} Paczynski, B.\ 1986, \apjl, 308, L43
\bibitem[Palmer et al.(2005)]{Palmer2005} Palmer, D.~M., Barthelmy, S., Gehrels, N., et al.\ 2005, \nat, 434, 1107
\bibitem[Pence(1980)]{Pence1980} Pence, W.~D.\ 1980, \apj, 239, 54
\bibitem[Pietrzy{\'n}ski et al.(2013)]{Pietrzy2013} Pietrzy{\'n}ski, G., Graczyk, D., Gieren, W., et al.\ 2013, \nat, 495, 76
\bibitem[Pozanenko et al.(2020)]{Pozanenko2020} Pozanenko, A., Minaev, P., Chelovekov, I., et al.\ 2020, GRB Coordinates Network, Circular Service, No. 27627, 27627
\bibitem[Preece et al. (1998)] {Preece1998}Preece, R. D., Pendleton, G. N.; Briggs, M. S.; et al., \apj,   496,  849
\bibitem[Rekola et al.(2005)]{Rekola2005} Rekola, R., Richer, M.~G., McCall, M.~L., et al.\ 2005, \mnras, 361, 330
\bibitem[Sari(1997)]{Sari1997} Sari, R.\ 1997, \apjl, 489, L37
\bibitem[Svinkin et al.(2015)]{Svinkin2015} Svinkin, D.~S., Hurley, K., Aptekar, R.~L., et al.\ 2015, \mnras, 447, 1028
\bibitem[Svinkin et al.(2020)]{Svinkin2020} Svinkin, D., Golenetskii, S., Aptekar, R., et al.\ 2020, GRB Coordinates Network, Circular Service, No. 27595, 27595
\bibitem[Tanaka et al.(2007)]{Tanaka2007} Tanaka, Y.~T., Terasawa, T., Kawai, N., et al.\ 2007, \apjl, 665, L55
\bibitem[Thompson \& Duncan(1995)]{Thompson1995} Thompson, C. \& Duncan, R.~C.\ 1995, \mnras, 275, 255
\bibitem[Thompson \& Duncan(2001)]{Thompson2001} Thompson, C. \& Duncan, R.~C.\ 2001, \apj, 561, 980
\bibitem[Vilardell et al.(2010)]{Vilardell2010} Vilardell, F., Ribas, I., Jordi, C., et al.\ 2010, \aap, 509, A70
\bibitem[Yang et al.(2020)]{Yang2020} Yang, J., Chand, V., Zhang, B.-B., et al.,  2020, ApJ, 899, 106.
\bibitem[Zhang et al.(2018)]{Zhang2018} Zhang, B.-B., Zhang, B., Sun, H., et al.\ 2018, Nature Communications, 9, 447





\end{thebibliography}
\end{document}